\documentclass[10pt,manuscript,nonacm=true,sigconf]{acmart}

\usepackage{booktabs} 

\usepackage{algorithmic}
\usepackage{graphicx}
\usepackage{textcomp}
\usepackage{xcolor}
\def\BibTeX{{\rm B\kern-.05em{\sc i\kern-.025em b}\kern-.08em
    T\kern-.1667em\lower.7ex\hbox{E}\kern-.125emX}}
\usepackage{graphicx}
\usepackage{multirow}
\usepackage{hyperref}
\usepackage{verbatim}
\usepackage{listings}
\usepackage{caption}
\usepackage{colortbl,xcolor}
\usepackage[normalem]{ulem}
\usepackage{footnote}
\usepackage{perpage}
\MakePerPage{footnote}
\usepackage{listings}
\definecolor{mygreen}{rgb}{0,0.6,0}
\definecolor{mygray}{rgb}{0.5,0.5,0.5}
\definecolor{mymauve}{rgb}{0.58,0,0.82}
\usepackage[T1]{fontenc}  
\usepackage{textcomp}     
\usepackage{lipsum}

\newcommand{\ssh}{\texttt{ssh}}
\usepackage{enumitem}
\newlist{todolist}{itemize}{2}
\setlist[todolist]{label=$\square$}

\begin{document}
\title{Cybersecurity Training for Users of Remote Computing}

\author{Marcelo Ponce}
	\email{m.ponce@utoronto.ca}
	\affiliation{
		\institution{Department of Computer and Mathematical Sciences, University of Toronto Scarborough}
		\streetaddress{1265 Military Trail}
		\city{Toronto}
		\state{Ontario}
		\postcode{M1C 1A4}
		\country{Canada}
}

\author{Ramses van Zon} \email{rzon@scinet.utoronto.ca}
\affiliation{%
  \institution{SciNet HPC Consortium, University of Toronto}
  \streetaddress{661 University Ave, suite 1140}
  \city{Toronto}
  \state{Ontario}
  \postcode{M5G 1M1}
  \country{Canada}
}



\begin{abstract}
End users of remote computing systems are frequently not aware of
basic ways in which they could enhance protection against
cyber-threats and attacks.  In this paper, we discuss specific
techniques to help and train users to improve cybersecurity when using
such systems.  To explain the rationale behind these techniques, we go
into some depth explaining possible threats in the context of using 
remote, shared computing resources.  Although some of the details
of these prescriptions and recommendations apply to specific use cases
when connecting to remote servers, such as a supercomputer, cluster, or
Linux workstation, the main concepts and ideas can be applied to a
wider spectrum of cases.
\end{abstract}


\maketitle


\section{Introduction}
\label{sec:intro}

In the last decade, scientific computing, or advanced
research computing, has seen a sharp increase in the utilization of computational
resources outside of traditional disciplines like the physical
sciences and engineering \cite{Ponce_2019,van_Zon_2019}.
Nowadays, computational resources are shared with disciplines requiring novel
approaches to problems and questions such as digesting and
analyzing copious amount of data, simulating models for predicting possible
outcomes, and statistically evaluating the support for empirical conjectures.
Disciplines such as medical sciences, biological sciences, bioinformatics,
machine learning and artificial
intelligence have emerged as the heavy users of 
digital research infrastructures, from computations to storage allocations.

Not so surprisingly, at the genesis of these
emerging computational fields, their practitioners were not
necessarily savvy or formally trained in technical areas such as
programming and high-performance computing.
Significant progress and effort in advancing computational knowledge
in these fields has been made, although some
areas still remain to be improved.
In particular, cybersecurity is one area in which not just new
scientific computing practitioners but also more experienced ones would benefit from
more in-depth awareness.

Furthermore, newer fields such as medicine and biochemistry can bring
more sensitive data, such as that collected from individuals, than the
more traditional fields.  To keep up with a changing environment, such
as the increase in working from home, security requirements and best
practices keep evolving, and users from all fields will need updated
instructions and retraining.

The main goal of this paper is to show what security mechanisms and
best practices should be common knowledge for \textit{end-users} (as well as and \textit{the
  support organization}) when using remote resources
such as supercomputers or advanced research computing, based on
experiences in training users of the supercomputers at the SciNet HPC
Consortium at the University of Toronto \cite{Ponce_2019b}. 

General guidelines on how to remain safe online have been discussed and
summarized in multiple publications, e.g. \cite{10.1371/journal.pcbi.1008563}.
Similarly, good recommendations on how to strengthen and improve passwords are
presented and discussed in \cite{10.1145/3372297.3417882}.
However, it was not until very recently that even specialized organization, such as
the National Institute of Standards and Technology (NIST) formally
recognized and began a campain to address the issue of cyber-security
standardization in High-Performance Computing systems \cite{NIST_HPC-cybersec}.

A second goal for this paper is to be a practical reference for these
security techniques.
Security is always a moving goal, but we aim to present currently
available and appropriate security techniques.
We will focus on the following key elements:
        i) using authentication methods which are more robust and reliable
for connecting to remote resources than passwords;
        ii) concrete practical implementations to be
followed when remotely connecting to servers, clusters, supercomputers, or even
remote workstations from work, labs, or home;
        iii) concrete recommendations and tools for users to help
protect while working connected to remote systems.

This paper is organized as follows:
in Sec.~\ref{sec:cyberSec} we introduce the most relevant and important concepts of cybersecurity,
in Sec.~\ref{sec:remotesharedSec} we explain the characteristics of remote
shared resources,
in Sec.~\ref{sec:attacks} we briefly present the most common type of
cyber-attacks currently known.
Sec.~\ref{sec:training} summarizes SciNet's current training program.  To
motivate this program and as a reference for future training material,
in Sec.~\ref{sec:CSBP} we describe guidelines to basic cybersecurity
best practices to mitigate some of the main issues presented in the
Sec.~\ref{sec:attacks}, and what role training should play in implementing
these practices.
Sec.~\ref{sec:conc} finishes with conclusions.

Additionally, to be able to reflect updates and addendums to best
practices, we have created a public accessible repository containing these recommendations, as well as further details
and more technical aspects of some of the strategies described in this paper.
The location of this repository is 
\url{https://github.com/cybersec-BestPractices/cybersec-RemoteComputing}.

\section{Security Context}

\subsection{Cybersecurity}
\label{sec:cyberSec}

The term cybersecurity refers to the different techniques, strategies
and methods that can be applied or employed to protect assets and resources
against different types of attacks.
In particular the ``cyber'' aspect arises from the fact that the
assets are identified as ``electronic'' or ``digital''. In many cases,
this is data and information stored in digital formats in computer servers and
remote machines.

Cyber-attacks, then, are the activities identified as threatening, disrupting, or attempting to gain access to the information illegally, i.e. without authorization.

Many strategies have been developed, and continue to be developed, to protect the confidentiality (i.e. only authorized parties can view the data), integrity (i.e. the data is not unexpectedly modified) and availability (i.e. the data or system is accessible) of digital data or systems.
At the same time, cyber-threats continue to grow at a substantial and
significant pace
\cite{verizon2021report,beek2019mcafee,internet2018cyber,doi:10.1016/S1353-4858(20)30079-9}, both
in complexity and number.

Critical to understanding attacks and
protections against them, is identifying
\textbf{vulnerabilities}. Vulnerabilities constitute weaknesses or flaws in systems.
Such vulnerabilities can originate from poor designs,  oversights
of some parts in the systems, uncorrected bugs, or unforeseen
use cases.

No system can be guaranteed to be 100\% attack-proof,
and very stringent approaches could come at the the expense of
usability.  Because of these reasons, the risks and severity of a
breach must be weighed against the cost of protective measures and the impact to
usability\footnote{This is usually done employing the so-called \textit{cybersecurity matrix} -- see our repository for more details about cyber-security.}.
For remote shared computing systems meant for academic research, this balance
will have a different outcome than for e.g. an online banking site.

\subsection{Using a Remote Shared Computing Resource}
\label{sec:remotesharedSec}

We need to discuss a few characteristics of Advanced Research
Computing (ARC), or High Performance Computing (HPC), facilities\footnote{We should also note that for the scope of this paper, ARC and HPC systems will be considered equivalent.}
before we can introduce what specific threats mean in that environment.
We should also note that for the goal of this paper, we will consider ARC and HPC
systems as 

First of all, access to such systems is usually remote, which means a connection needs to made
over the internet. The internet is the paradigmatic
example of connectivity. Its confluence of heterogeneous systems also
opens up vulnerabilities.
At its origin, the internet was a group of mutually
trusting entities attached to a transparent network, and not designed with much security in mind.
One's default attitude should be to not trust what is on the internet.
This particularly applies to websites, where directly or indirectly visiting
a fake websites may result in direct exposure and potential attacks to the connecting devices.

The most vulnerable element here is the user and their behaviour.  Beginning
users of HPC systems may be aware of security concerns using web-based
authentication and access, but access methods to remote HPC resources
are often unfamiliar. Without training
they will not know best practices nor how to remain vigilant in using
these systems.

One of the most common methods of connection is via ssh, which stands
for ``secure shell''. In this context, let us call the computer from which the users logs in the ``local''
computer, and the HPC facility the ``remote server''.  To start the
connection from the local computer to the remote
server requires the user to authenticate.  For a long time, authentication would
be based on a username and password, but that is no longer a best
practice.  While the connection is active, data
flows between the user's computer and the remote facility.  Ssh
connections provide encryption of this data flow.

Additional features of ssh that are common, but have security
implications, are support for graphics windows using X forwarding, port
forwarding for reaching hosts inside the remote facility, and key-agent
forwarding to facilitate authentication to other facilities from the
first remote facility.

The remote setup also means that one should be concerned with the
security of both the local computer and the remote server.  If the local
computer is itself a shared computer, for instance, one that is
present in a research laboratory, that can pose additional concerns.

The remote server is a shared system, usually running a flavor
of GNU/Linux or UNIX as the operating system.  Such operating systems
make a distinction between privileged users and regular users, sort users in
groups, and maintains ownership and group membership of files and
running programs, that can and should be used to control access to
files and commands, as there are typically many (regular) users logged
in.

\subsection{Types of Cyberattacks and Cybersecurity Threats}
\label{sec:attacks}

There are many types of cyberattacks, ranging 
from specially targeted and designed attacks to more generic and
opportunistic ones, such as the so-called \textit{zero-day exploits},
in which attackers take advantage of a vulnerability for which a patch has not been developed yet.

In this section, we will review some of the most common types of attacks and the impact they could have on users and services.
While they could utilize many different approaches, one could classify them in two basic categories:
	one where the attack's goal is to impact availability (e.g. by bringing
down a particular functionality in a system);
	another class of attacks where the goal is to impact confidentiality and/or integrity (e.g. attempt to gain access to
unauthorized resources, such as systems, privileges or/and
users accounts).
The techniques, tools and best practices to mitigate these different types of
attacks, both by the ARC service providers as well as their user, will
be discussed in section \ref{sec:CSBP}.

We can distinguish several objectives of cybersecurity attacks:
\begin{enumerate}
\item[a.] Get past authorization to get access to a system (``Hacking'');
\item[b.] Disable a service (e.g, by ``Denial of Service attack'');
\item[c.] Steal secure information (e.g., through ``Phishing'');
\item[d.] Install software on a system that can be used later for
  later attacks (e.g. ``Malware infection'').
\item[e.] Abuse resources (e.g, Cryptocurrency mining when this is
  against a  ``Terms-of-Use'' agreement)
\end{enumerate}
Cybersecurity attacks usually have several of these
objectives.  Most of these
attacks are crimes in many jurisdictions \cite{cybersecuritylaw2023}.

In a so-called \textit{brute force attack}, an entity 
will attempt to get access to a system by systematic attempts to guess user credentials to
authenticate in the targeted system, e.g. a username and password.
Nowadays, brute force attacks often rely on advanced tools to try many different passwords (not just 'guessing' which suggests a manual process, in which the attacker may know something about the victim).
Brute force attacks are still quite effective despite the existing controls to prevent them.
According to the 2017 Varonis' data breach report \cite{varonis-BFA}, `5\% of confirmed data breach incidents in 2017 stemmed from brute force attacks'.

Once access to a service has been gained, the consequent risks depends
on what authorization the user has whose credentials were obtained.
Regular users would only have access to their own data, or to any data
shared with them, and the impact of the breach could be limited to
just their account.  Administrators and staff may have elevated
access, and having their accounts hacked would be much more dangerous
and impact several users or even the whole system.  It should also be
noted that unintended security vulnerabilities in the software used in
a service or its operating system might make it possible for regular
users to gain administrative powers.

Another common type of attack is the so-called \textit{denial of service} (DoS),
in which an attacker would attempt to bring down --partially or completely-- a system
or network.
There are different methods in which this can be done. A common method is by
flooding with traffic a given system so that it saturates its resources
or even the bandwidth of the network.

Since frequent traffic from a single IP could easily raise flags and
be stopped, a more elaborate version of this type of attack involves
launching 
DoS attacks from various, distributed servers. This
case is referred to as a ``distributed''-DoS or DDoS.

\textit{Malware} is a general term used to refer to any type of malicious
computer programs.
There are different types of malware, among the most ``famous'' ones are:
viruses, worms, ransomware, Trojan horses, rootkits, etc.
Its goal can range from making a
system unresponsive, steal information (credentials, documents, etc.), "kidnap"
information, espionage, use a connection point to jump to another systems to
hide the trace of an attack, etc.

Malware can find their way onto a computer e.g. as part of other
software packages, which may have gotten installed as part of a
packages, or be installed unintentionally by cleverly disguised
website.   While they usually target users' personal computing
devices because they have administrative permissions on them, servers
can also be infected in the form, for example, of so-called root kits.

\textit{Sniffing, IP/DNS Spoofing, and Man-in-the-Middle attacks} 
are a collection of techniques aimed at getting information out of
the data that is transmitted.  Sniffing refers to collecting the
packets of information while it is transmitted through the network.
When using encryption,  the information in the packets themselves
can't be read, if the encryption is strong enough; this is why the
type of ssh key matters.

IP Spoofing refers to a technique where a malicious party attempts to inject
information into the network as if it came from other system, e.g.,
the actual remote server.  The objective of this attack would be to convince the user or other
systems that the message comes from a valid source and in this way establish
an exchange of information.  In this way, credentials or other
sensitive information could be obtained.
In combination with a DoS or DDoS attack, this attack can be used to
disguise and redirect network traffic to malicious and bogus sites.

Man-in-the-middle is a term used to describe a third party that is
attempting to eavesdrop 
or intercept information sent between the user and the remote system.
This can happen in different ways, for instance, at a physical level, where
a device or connection can be added to the main communication channel; or, at a 
``software'' level, where similarly a program can be employed to intercept and
steal the data shared within the communication.

One element which is often essential in cyberattacks
is the ``human factor'' which involves taking advantage of certain
characteristics in standardized human behavior by tricking people to
divulge sensitive or private information, in order to obtain access to systems or steal information.
This type of attack is commonly known as ``social
engineering''. The level of sophistication can vary from very generic to more targeted and specialized.

Phishing techniques are a subcategory of social engineering attacks. They relate to illegitimate emails attempting to
acquire \textit{sensitive data} by exploiting the victim's inexperience and trust. They are a very popular mean
to obtain credentials, and from there hack into an organization.

Any of these (illegal) cyber attacks  can be hard to
detect, and hard to fix once the damage is done. The best approach is
to try to make such attacks less likely to suceed. The best approach for
protection depends on
the type of attack.

\section{Security Training Program}
\label{sec:training}

The need for training for various forms of user training will be
explained in the next section, but for clarity, it is worth to point
what security-related training SciNet has offered to its users so far.  These courses can be found on https://education.scinet.utoronto.ca by searching for
the course codes given in the parentheses below.

\begin{small}
\begin{description}
  \item[Intro to SciNet, Niagara, and Mist (HPC105)]
\ \\Typically given as a single session of 90 minutes, this presents the
details of logging in into the Niagara and Mist clusters (including
using ssh and keys), available file storage, and creating and
submitting jobs to be the schedule.

\item[Intro to the Linux Shell (SCMP101)]
  \ \\This 3-hour workshop familiarizes new users with the Linux shell,
  which is the  main interface to our systems.

\item[Introduction to Supercomputing (HPC101)]
  \ \\Either given as one session of about 3 hours, or in 3 separate
  sessions, this workshop shows why (remote) clusters are used, as
  well as common ways people use it.

\item[Securing File Access Permissions on Linux (SCMP283)]  
  \ \\This workshop 
  is aimed to educate users about what permissions are, how to use
available tools to control access and sharing, and how to avoid common security
pitfalls.

\item[Introduction to Apptainer (SCMP161)]
  \ \\This workshop introduces users to Apptainer, a container solution,
  which could be used for software that
  requires a specific OS setup different from what the cluster uses,
  or to handle workflows with many files, or for enhanced security.

\item[Enable Your Research with Cybersecurity (SCMP183)]
\ \\A workshop of 4.5 hours given over the span of three days, that covers
various aspects of cybersecurity, cyberattack models, and best
practices. Also covers cybersecurity in the context of human research data and the Research Ethics Board. 
  
\item[Advanced Linux Command Line (SCMP271)]
\item[Bash command line with common idioms (SCMP281)]
  \ \\Both of these are workshops that allow users to further their Linux
  skills.  

\item[SSH Keys Drop-in Session (SCMP110)]
  \ \\In the ssh keys pilot
  (Oct 2021-Jan 2022) in which password authentication was replaced by
  ssh keys on
  SciNet systems, several drop-in sessions were held to help users set
  up their ssh keys.

\end{description}
\end{small}

In our cybersecurity training, as in most of our training, the
strategy is not to only provide general information, but to offer
courses that combine such information with concrete practical exercise either done
during the sessions or reviewed and graded with feedback afterwards.  In addition, we offer support for users
after these sessions.  This post-training support makes it much more
likely that the security practices are indeed followed, and also makes
it possible to enforce some practices.

For security training it is possible to assess its success by aggregating
how users use the system. For instance, during the ssh keys pilot, we
tracked the number of users using keys, and saw it steadily
rise as we gave a training session on the upcoming changes (see
Table \ref{tab:passwordusers}).

\begin{table}
  \begin{tabular}{|c|c|}\hline
    \textbf{Week ending on:}&\textbf{Users using keys}\\\hline
Oct 10, 2021 & 38\%\\
Oct 17, 2021 & 43\%\\
Oct 24, 2021 & 47\%\\
Oct 31, 201 & 50\%\\
Nov 7, 2021 & 59\%\\
Nov 14, 2021 & 62\%\\
Nov 21, 2021 & 67\%\\
Nov 28, 2021 & 76\%\\
Dec 6, 2021 & 76\%\\
Dec 12, 2021 & 86\% \\
Dec 19, 2021 & 82\% \\
Dec 26, 2021 & 79\% \\
Jan 2, 2022 & 80\% \\
Jan 9, 2022 & 78\% \\
Jan 16, 2022 &89\% \\
Jan 22, 2022 &100\% \\\hline   
  \end{tabular}
  \caption{Week-by-week progress in promoting ssh keys over password
    authentication to log into SciNet's Niagara cluster. Password
    access was disabled on January 22, 2022.\label{tab:passwordusers}}
\end{table}

While in-person training is often preferable so that issues on users' own
laptops can easily be addressed, nonetheless, because the systems hosted
and operated by SciNet are used by researchers
throughout Canada as well as their collaborators outside of Canada,
most of these training sessions were online.

SciNet's courses are open to all users of the Canadian national ARC
facilities as well as to members of Canadian academic institutions.
Beyond this target group, all materials of SciNet courses are freely
available online to anyone on
	\url{https://education.scinet.utoronto.ca}.

\section{Cybersecurity Best Practices}
\label{sec:CSBP}

There are several lines of defense that the provider of the ARC service
should establish proactively.  
\begin{itemize}
	\item Securing authentication:
		This involves verifying the identity of an entity, user, process, program, server, etc.

	\item Protecting authorization:
	Restricting access to certain users, based on their identities and qualifications serves the purposes of preserving the privacy and confidentiality of the data; examples of implementations are role-based access controls.

	\item Using encryption: Particularly when transferring data,
          but sometimes also required for data as it is stored in the
          ARC center.

	\item Integrity checks:
	This is quite relevant in order to guarantee that the data
        that was sent has not been tampered with and is
        trustworthy. On an operating system level, it is important
        to check that
        no system executable are changed and replaced by malware.
	There are different techniques to implement integrity checks,
        some of the most common ones include digital signatures or
        checksum calculations.

	\item Network filtering:
	This is to limit traffic coming in and out of a system or
        security perimeter (e.g. firewalls). Implementations can be
        done at the software or hardware level.
\end{itemize}

While the aforementioned approaches are put in place by the ARC
center, many users may need to know about them to understand if they
are allowed to use that system for their data.  In traditional ARC
systems, the responsibility for data access control was often left to the users.
For some specific types of data, stricter guarantees are needed that require more
measure from the ARC center, and various auditable certification levels
exist \cite{CMMC}.

In addition to measures put in place by the ARC center, users also have a responsibility to protect against
cyberattacks, because attacks often do not start on the remote system,
but on the end-users local computer.   We will present several
specific strategies for end-users below.

\subsection{Software Updates}
The most basic and immediate way to improve the security of a user's
local computer is to keep
that system's operating system (OS) and programs update to date.
Many times attackers will take advantage of systems which are not up-to-date with
the latest releases or security patches for the system, and gain access by exploiting
\textit{vulnerabilities} that could have been mitigated by a simple OS or
application update. 
Should the end user's workstation be compromised because their machine was not patched in a timely fashion, it could also lead to the compromise of the remote computing system they are connecting to.  Therefore, by keeping their systems (desktop, laptop, etc.) up to date, the end user also helps protect remote computing systems.

The best practice for users, as well as for center staff the workstations, is quite simple, keep your systems up to date!

\subsection{Antivirus \& Malware}
Computer viruses and malware are
potential high-risk entry vectors to local computers and through
those, to our machines. 
As such, having \textit{antivirus} software installed and running on
local computers is critical.
Users should be encouraged to check with their university IT department or
library, which usually provide licenses for students and staff to get
antivirus products.

Antivirus software can detect malware signature's from a database
which is updated on a regular basis, but have started to use machine
learning techniques to identify unknown or file-less malware which was
not previously detected.

\subsection{Authentication enhancements with \ssh}

A very common protocol for connecting to a remote server or system is \ssh.
\ssh~stands for \textbf{s}ecure \textbf{sh}ell. It creates
an encrypted channel between the client (user trying to connect) and the server
(system where the user wants to connect). While \ssh~offers a secure way to connect between computers,
it can be vulnerable to some of the attacks described in Sec.~\ref{sec:attacks}, such as man-in-the-middle attacks or brute-force attacks.
We will not go into the details of how \ssh~ creates this secure communication
channel but we will instead focus on the mechanism to authenticate
the user in the remote system, as it plays a key role in mitigating the risk of attacks against ssh.

\subsubsection{Passwords}

At the moment, the most common way of authenticating in ARC systems is by using a
username and password.
Passwords may give users the illusion of protection, but they are one
of the least secure authentication methods. 
Passwords can be compromised, can be weak, can be stolen, and of course
are in most cases chosen by humans -- who arguably can be considered the
weakest element in the cybersecurity chain.

When there are better alternatives, as the ones mentioned below, they
should be used. But many authentication methods still rely
on passwords utilization.
Whenever this is the case, an additional tool to consider to use is a 
password manager that stores passwords encrypted.
In addition, password managers can help to organize and even validate or
check the strength and integrity of passwords.
Depending on the OS there are different options available, a couple of open
source options are:
\texttt{KeePassXC} (\url{https://keepassxc.org})
and 
\texttt{bitwarden} (\url{https://bitwarden.com}).

The main risk with password authentication is the ability for an
attacker to obtain these credentials. Since the password needs to be
transmitted to the remote site, there is a possibility that it may be intercepted.

\subsubsection{\ssh~keys}

A generally more secure and efficient way to authenticate users with a
remote system that does not suffer from the vulnerability that
passwords have, is to
use \textit{keys}.

The authentication via keys leverages asymmetric encryption. It involves two keys which are part
of a key pair: one \textit{private key} which must be kept secure, and one \textit{public key}, 
which can be distributed. The public key is used to encrypt data, which can be decrypted with the 
corresponding private key. After the establishment of the ssh connection, the user's authentication 
occurs. The remote server sends an encrypted challenge request, encrypted with the public key, to the client. 
The client then decrypts the challenge request with the private key, and sends it 
back to the remote server. Then, the remote server compares the two pieces of information 
(the challenge request, versus the challenge response by the client), and if they match, 
the authentication of the user via ssh keys is successful. It is important to note that these 
steps are transparent to the user. Also, the private key never leaves the client, making this 
method of authentication more secure than the authentication with password.

The process of starting to use an asymmetric keys pair for \ssh~ can
be summarized as follows:

\begin{enumerate}
        \item Create an SSH key pair on your \textbf{local} machine --
        on a Linux, Mac OS or even Windows using MobaXterm or Linux-subsystem terminal,
	this can be done using the following command:
\begin{lstlisting}[language=sh,numbers=none,firstline=2,lastline=2]
ssh-keygen -t ed25519
\end{lstlisting}

	When this command is executed, it will prompt for the \textit{location} where the keys are going to be placed and for a \textit{passphrase} to associate to the keys.
The passphrase is like a password local to your computer; its purpose is to encrypt the private key to better protect it against potential theft.
After these two pieces of information are entered, the command will create a pair of files to be located at the location specified previously --its default location would be  \texttt{\$HOME/.ssh/}, where \texttt{\$HOME} represents the user's home-directory--.
	If no further details are given, the files are by default named as \texttt{id\_ed25519} and \texttt{id\_ed25519.pub}, representing the private and public keys respectively.
	The \texttt{-t ed25519} used in the \texttt{ssh-keygen} command represents the type of algorithm used to generate the encrypted keys and this one in particular is one of the recommended standards to be used nowadays.	

        \item The next step is to transfer the file with the \textbf{public} key to the remote server/machine.
        Different remote facilities will have different mechanisms for
        this.  On some, one can use the following command from the
        local computer:

\begin{lstlisting}[language=sh,numbers=none,firstline=5,lastline=5]
ssh-copy-id  -i $HOME/.ssh/id_ed25519.pub  USERNAME@remote.system.ip
\end{lstlisting}

		where the \texttt{-i} flag indicates the public key
		file (in the example the default one located
		at \texttt{\$HOME/.ssh/id\_ed25519.pub}) to be copied
		over the remote system --\texttt{remote.system.ip}--
		to which the user \texttt{USERNAME} would like connect
		to. 		At this point the user will still be asked to authenticate itself, by entering its username/password combination.

                On other systems, there may be a web interface that allows
		users to upload the public keys.
                That site itself may use passwords, perhaps combined
		with MFA (see further down).

          \item After these two steps, unless the default location was
          used for storing the private key, any time one uses the ssh
          command, it must be told where to find this key with
          the \texttt{-i} flags.
\end{enumerate}

At this point some few observations should be done:
\begin{itemize}
	\item Having a combination of private/public-keys guarantees that only a machine where the private key can be found can connect to the remote location where the public key resides. Hence why is critical that the private key \textbf{never} leaves the machine where the keys were generated, this also includes not copying them to any other machines.
	
	\item There should be an unique set of keys per machine.
		 In other words, if a user owns a laptop, a desktop
		 and a workstation, the procedure described above
		 should be repeated independently in each of these
		 devices.
                 One may also want to generate separate keys for each
		remote system, if one wants each trusted relation to
		 have a unique key pair.

	\item The private key should always be protected with a passphrase, ie. never leave an empty passphrase!
When not specifying a passphrase, the private key remains
unencrypted. If someone gains access or takes control of the local
device, they will be able to connect to the remote system.
\end{itemize}

Many more details can be added to the process of keys generation.
These details are presented and discussed in 
our associated repository,
\url{https://github.com/cybersec-BestPractices/cybersec-RemoteComputing}.

We should note that \ssh~ keys themselves are also prone to brute force
attacks if the length of the key is too short and/or the algorithms used are deprecated.
The National Institute of Standards and Technology (NIST) in the US,
has developed a series of reports \cite{barker2020recommendation,barker2020cryptokeygen}
describing the recommended standards to use for keys encryption algorithms along with
keys' length (see Table~\ref{table:NIST_ssh_std}). 
It is advisable to stick to these NIST standards in order to minimize
the risk of brute force attacks.

\begin{table}[h]
\begin{tabular}{| p{.25\columnwidth} | p{.2\columnwidth} | p{.4\columnwidth} |}
	\hline
		Encryption Algorithm	&	Key length	&	key generation command
	\\
	\hline\hline
		ECDSA, EdDSA, DH, MQV	&	$224-255$ (and above)	&
			\begin{minipage}{.4\columnwidth}
\begin{lstlisting}[language=sh,numbers=none,
				frame=none, firstline=1,lastline=1]
ssh-keygen -t ed25519
\end{lstlisting}
			\end{minipage}
	\\
	\hline
		RSA			&	$2048$ (or above)	&
			\begin{minipage}{.4\columnwidth}
\begin{lstlisting}[language=sh,numbers=none, frame=none, firstline=3,lastline=3]
ssh-keygen -t rsa -b 4096
\end{lstlisting}
			\end{minipage}
	\\
	\hline
\end{tabular}
\caption{NIST's standard recommendations for \ssh~ keys encryption algorithms \cite{barker2020recommendation,barker2020cryptokeygen}.}
\label{table:NIST_ssh_std}
\end{table}

Theoretically, \textit{quantum computers} would be capable of breaking the 
cryptographic algorithms that are used in ssh.  Although it is not clear how soon quantum computers will
be powerful enough to do so, in response to the developments in quantum
computing,
NIST has already began the preparation for the so-called
\textit{``Post-Quantum Cryptography Standardization Process''} \cite{NIST_PQC}.


As was mentioned above, the \texttt{ssh-copy-id} may not work on some
systems, particularly on those sites that have further enhanced the ssh-key
mechanics with a \textit{centralized SSH-keys} database.  The CCDB of
the Digital Research Alliance of Canada provides such a capability for
the national ARC systems in
Canada.\footnote{\url{https://docs.alliancecan.ca/wiki/SSH_Keys}}.
Users can upload their public keys which will then be used in multiple
remote systems. This will happen transparently for the users, as the
underlying infrastructure will take care of propagating the
information across the different systems.

On one of these national systems, the Niagara cluseter at SciNet,
centralized ssh keys are the only mechanism of
authentication, 
Following the pilot of about four months (see Sec.~\ref{sec:training}).

Because there is a learning curve to using ssh keys, and because the
methods of setting it up depends on the local operating system and ssh
clients, a combination of incentives (such as brownout periods --i.e. periods
with a reduction or restriction in how users could connect to the system--
using password authentication) with training and drop-in sessions has helped
deliver a smooth adoption on Niagara.


\subsection{More secure authentication with MFA}

Multi-factor authentication (MFA) is a widely utilized 
security measure in various technological domains, aimed at ensuring
additional layers of authentication.  Users may be familiar with
implementation of MFA in mobile devices, where
biometric factors such as facial recognition, iris scanning, and
fingerprint sensors are utilized to authenticate users.  But MFA is a
more general technique that enhances
secure authorization by require multiple separate pieces of
authentication, called factors.

The main benefit in security is to go from a single factor to two
factors, i.e., to have two-factor
authentication, also known as second-factor authentication (2FA).  The
benefit of adding more authentication factors to the same service tends to be marginal.

There are multiple and diverse MFA mechanisms and implementations,
some of which are open-source and free, while others are commercial.
Among the more popular choices are time-based approaches which
generate a one-time-password (OTP) to use when authenticating. Such a
code can only be used once and for a short and specific period of
time.  This concept is also used in commercial services, like
telephone companies or financial institutions, to validate their users
credentials by contacting them on their phones as a second way to
authenticate their identities.  It is also possible to use this form
of MFA in combination with ssh using the open-source
GoogleAuthenticator
(\url{https://github.com/google/google-authenticator}), or PrivacyIdea
(\url{https://www.privacyidea.org/}) which is another free open source
initiative.  Alternatively, there are commercial solutions, such as
Duo. These commercial solutions offer support for pushing an
authorization request to a user's cell phone or accepting
hardware devices like YubiKeys.

Academic institutions are increasing adopting MFA as well for
authenticating.  Because many ARC centers like SciNet serve users from
several institutions, they require their own MFA implementation, but
if it is the same solution, users may be able to reuse the same app.
SciNet has used Google Authenticator as an optional MFA method for
users (and a mandatory one for its staff) since June 2020.  After
having taken part in
several pilot projects for MFA across the Canadian national ARC
systems under the umbrella of the Digital Research Alliance of
Canada, has transitioned to using Duo in May 2023.

While users could add MFA for their computers, e.g, using one of the
open source solutions, if their computer is not accessible from the
wider internet, the security benefit seems small.  But if MFA in
connecting to their ARC facility is
available, they should be encourage or required to do so.

\subsection{VPN}
An additional layer of protection that users can add when working or connecting
remotely is to use a \textit{Virtual Private Network}.
The main objective of this type of technologies is to extend the domain of
a private, secure, controlled networks beyond the physical limits that would
usually define such a network.
VPN offers a secure, encrypted connection over a shared network.
A typical example is a VPN offered by an academic institution,
which would allow their students and personnel to remotely connect to its
network as if they were on campus.
This offers multiple advantages, such as having an IP address assigned within
the domain or range of IP addresses within the academic institution, additional
protection against undesirable Internet traffic or malicious agents.
It is also possible to engage with private providers of VPN services, although
we would encourage users to inquire with their IT departments and libraries
within their corresponding academic institutions is such a service is available.

VPNs can also be a means to mitigate the security risks of other methods of connecting. For instance, the employment of Virtual Network Connections (VNC) is a common practice when
working remotely. It offers the remote user a great deal of flexibility and much
more responsiveness in what it refers to graphical interfaces, than other possible
counterparts like X-forwarding over \ssh~ connections.
However there are a couple of elements that are usually considered risky in terms
of security: many VNC systems allow for users to connect without the use of a password,
which needless to say is a highly discourage practice!
Secondly, VNC works by opening connections through a given port in a server,
these connections --which by design are resilient-- should be tear down when
not used to reduce the chance of ports swiping by a malicious party.
In many cases, specially in supercomputer centers, where resources may not be directly exposed
to the Internet, the best way to reach a service like this is by \textit{tunneling} through 
the so-called login nodes.

\subsection{Further Protection against cyber-attacks}

As protection against brute force attacks, 
ARC centers should have controls in place
that detect repeated authentication failure attempts, resulting in
the application of a banning policy. 
For instance, some approaches will ban users from accessing the system for a
period of time, or lock their account and request a mechanism to unlock it using
another mean to authenticate the user, eg. email or SMS.  

Limiting the number of connections per minute mitigates much of the
brute force attacks, but there are ways that end users can further
mitigate the risk of brute force attacks by choosing longer usernames
(the number of possibilities to try for short usernames is small)
Similarly, users should avoid having simple, repeated/reused, or short passwords, or even better
avoid using passwords at all by substituting them with ssh keys.  Additionally,  private keys should be protected
with a strong passphrase, and never leave the local computer.  Note
that, ideally, all this
should be taught even before the user accesses the remote
server for the first time.

To mitigate
denial-of-service 
attacks, frequent subsequent
authentications, successful or not, will trigger
a banning policy on the originating IP.  It is important to inform
users of this limitation, and to discourage connecting many times per
second or transferring many separate files instead of combining the
files into one zip or tar archive files and transfer that.  The
IP-based banning (even if temporary) can be quite disruptive for
research labs where the local computer is shared, or in which the
local computers share a single outbound IP.

ARC center can be expected to mitigate the risk of malware with configuration
management, restricting root access, a rootkit scanner, etc.
End users of remote computing systems have a role to play in
protecting themselves against malware by running antivirus software
and malware scanners, even on Macs and Linux computers.
It is advisable for users to have separate machines for private and
research, if they can.  On their local computer, encourage them not to
blindly click yes on popup windows, and to look at all warnings,
errors and messages. 

There are different ways for ARC centers to mitigate man-in-the-middle
attacks, such as encryption, implementation of integrity checks to
verify that the data has not been manipulated, etc. While most
controls to protect against such attacks fall under the responsibility
of the administrators of the remote computing systems, here again the
end user has a role to play.

An example of this, is \ssh~ trying to warn users about possible MITM attacks
when checking for the "fingerprint" codes of known systems when these change
in comparison to previous connections or sessions. The end user should be particularly vigilant if such warnings are displayed in their terminal. They should be aware of these fingerprint codes and their actual
values in order to verify the authenticity of the servers one would be connecting to. 

Thus, ARC centers should advertise the fingerprints. 

One of the best ways to prevent phishing attempts is to educate users! \cite{verizon2021report}.
Check with your university's IT department or library, in general these
departments have resources available to instruct and educate users in how to
recognize phishing attempts.
For example, as many other institutions, the University of Toronto has collected
some examples at 
\url{https://securitymatters.utoronto.ca/category/phish-bowl/}.

In many cases, some attacks could happen without the victim
being even aware of it. A typical example is users whose accounts
have been compromised and are just being used as "trampolines" to jump
to other systems.
In other cases, attackers may just want to gain access to computational resources
in order to have more compute power at their disposition and for instance,
mine crypto-currencies. These examples have been detected multiple times
in different supercomputer centers or systems which offer substantial amounts
of computational resources \cite{bbcnewscrypto}.
Some simple strategies in order to mitigate this
unnoticed-driven and subtle abuse of users' accounts, is for systems to inform the
users about details on their connections; e.g. when and from
where were their last connections to the system, or even more sophisticated ones
such as, keeping track of the usual pattern of connections for users (e.g. IP,
geographical location, etc.). The end users should pay attention to these details (which are generally provided at the beginning of the session, in the banner message) and confirm that the activities belong to them. When certain irregularities are detected, they should report the anomaly to the administrators of the remote computing systems.

\subsection{Containerized Solutions}

\textit{Containers} have been quickly gaining popularity in the last few years,
as their approach offer a simple and robust solution to installing software
with multiple or complex dependencies.
Containers are also used to isolate resources and services, and in particular can
be a great solution to mitigate the escalation of security risks by differentiating
and separating into multiple containers.
For instance, in the case of an application deploying an attack within a container, this can
still shield it from the host; similarly an attack on the host could be shielded from
reaching hosted containers; as well as inter-containers attacks.
Among the most popular solutions are \textit{apptainer}\footnote{Previously known as \textit{singularity}.} and \textit{docker} containers.
Apptainer containers are usually recommended over docker ones due to security concerns --
mainly due to the fact that docker images require access to root privileges presenting
a potential high-risk liability.
Similarly users employing already prepared images should be aware of the risk of
utilizing ones coming from untrusted sources.

\section{Conclusions}
\label{sec:conc}

In this paper, we have presented a basic overview of the most typical forms of
cyber-threats in using remote computing facilities.
We discussed several useful techniques that end-users can leverage to mitigate some of these attacks.
Some of these techniques are well-known and commonly used by professionals in
the disciplines of computer science and systems administration.
However, many end-users with backgrounds in diverse disciplines may
need training in these (for them)
novel techniques.  Having put them in the context of the
risks and impacts, we believe will increase the
general awareness and at the end benefit the whole community of remote-systems users.

Users of cloud services should also be concerned about security and privacy risks.
Although this paper focused on using traditional ARC clusters, at the very fundamental level, all what is discussed in this paper and the
techniques presented and recommended here will still be applicable to this
type of systems too.
Remote connectivity using ssh and its improved forms, such as keys and MFA,
should be used, as well as any other form of enhanced connectivity.
Nevertheless there are a few elements that may be different from our previous
discussion. For instance, if the user is responsible for deploying, installing,
configuring, administrating and maintaining its own machines and environments in
the cloud infrastructure, then special considerations should be given to
the OS installation, permissions and privileges, allowed services running in the
remote machine, open ports, etc.
If the cloud service will be used as a sort of web-portal or gateway, additional
attention should be paid to web services running on the machine, tight all
possible access points and methods, as well as, being compliant with
certificates and protocols standards.
It is always recommendable that if the end-user is not familiar with this type of
configurations, to request support from specialized personnel such as system administrators
or technical support to check on all of these.

It is critical when using shared resources and accessing them remotely, to
realize that the system as a whole is as weak as its weakest element.
Hence why considering the implementation of the combined techniques and
strategies presented here is highly recommended to improve the overall
cyber-security posture of the remote system and local users connecting to it.



As a final remark, we have created a publicly accessible repository,
	\url{https://github.com/cybersec-BestPractices/cybersec-RemoteComputing}
where we aggregate and present most of the best practices, concepts and
implementation details presented in this work.

We decided to present this information in this way, so that it can be updated
as technology, trends and threats change and advance.

At the same time, we allow users to use this as a consolidated reference,
contribute, keep track of changes.
We additionally enable issues requests for users and readers to ask questions or make comments.
Similarly we have enabled the wiki feature to allow for users' contributions --
which of course will be curated by the authors and collaborators.


\begin{acks}
We would like to thank the SciNet team for their involvement in the
cybersecurity program, and in particular Rapha\"elle Gauriau for
promoting cybersecurity and cybersecurity awareness at SciNet. We
would also like to thank  Michael Nolta for helping
to collect the numbers in table \ref{tab:passwordusers}.

The SciNet HPC Consortium is funded by Innovation, Science and
Economic Development Canada; the Digital Research Alliance of Canada;
the Ontario Research Fund: Research Excellence; and the University of
Toronto.
\end{acks}


\bibliographystyle{ACM-Reference-Format}
\bibliography{references.bib}


\end{document}